# Seashell-inspired polarization-sensitive tonotopic metasensor

Y. Liu,[1] V.F. Dal Poggetto,[2] A.S. Gliozzi[1], N.M. Pugno[2,3], F. Bosia,[1] and M. Tortello[1,a)]

[1] *Dipartimento di Scienza Applicata e Tecnologia (DISAT), Politecnico di Torino, Torino, 10129 Italy*

[2] *Laboratory for BioiBionspired, Bionic, Nano, Meta, Materials & Mechanics, Dipartimento di Ingegneria Civile, Ambientale e Meccanica, Università di Trento, Trento, 38123 Italy*

[3]*School of Engineering and Materials Science, Queen Mary University of London, London, UK*

Bioinspiration has widely been demonstrated to be a powerful approach for the design of innovative structures and devices. Recently, this concept has been extended to the field of elasticity, dynamics, and metamaterials. In this paper, we propose a seashell-inspired metasensor that can simultaneously perform spatial frequency mapping and act as a polarizer. The structure emerges from a universal parametric design that encompasses diverse spiral geometries with varying circular cross-sections and curvature radii, all leading to tonotopic behavior. Adoption of an optimization process leads to a planar geometry that enables to simultaneously achieve tonotopy for orthogonally-polarized modes, leading to the possibility to control polarization as well as the spatial distribution of frequency maxima along the spiral axis. We demonstrate the versatility of the device and discuss the possible applications in the field of acoustics and sensing.

## I. INTRODUCTION

It is widely accepted that over millions of years of evolution, Nature has generated a wide variety of biological systems exhibiting exceptional mechanical, optical, and thermal properties[1–5]. Man has therefore regarded these systems with great interest, investigating their fundamental working principles also in view of the realization of novel materials and devices in many different fields[6]. From the point of view of biological and bioinspired materials, most works in literature are related to the investigation of static or quasi-static properties[7]. There are, however, several works addressing different relevant dynamical frequency-dependent features [8–11].

Seashells have been extensively studied due to their hierarchical microstructure and the related quasi-static properties[12,13] designed to ensure mechanical protection from impacts and predator attacks[14,15]. Recently, the mechanical properties of the *Turritella terebra* and *Turritellinella tricarinata* seashells have also been addressed and characterized from the dynamical point of view [11], showing that the overall shell structure and its shape may also play a role in impact attenuation and vibration damping. Many different seashell structures share common geometrical traits and can be described by similar parametric equations[16,17], yielding, for instance, both planospiral or conispiral structures with the tuning of a few key parameters. The variable cross section of shells, which may feature an elliptical shape, defines their structure by describing a logarithmic spiral central line. While doing so, the cross section reduces in size and, in some cases, can also displace along the axis of the shell, giving rise to a conispiral shape, as it is the case of the *Turritella* (Figure 1(a), (b) and (c)). To obtain the real shell structure through these approximating mathematical models, internal intersections should also be removed, following for example the procedure reported in Ref. [17]. In this work, the authors used a numerical method to find and remove the self-intersecting regions, and finally re-meshing the regions associated with the sutures of the previously intersecting parts. The same type of logarithmic spiraling central line and grading of geometrical properties is also present in other natural systems, such as the mammalian cochlea[18,19], which exhibits a tonotopic behavior, i.e., presents a structure which enables the detection of exciting sound waves based on their frequency content due to specific regions that convert sound energy into neural impulses depending on the different locations of excitation[20,21]. Therefore, tonotopy allows different frequencies of input signals to be univocally spatially mapped at different locations on the cochlea basilar membrane, stimulating the appropriate motion receptor cells[22]. This behavior can be mimicked by metamaterials using locally resonant structures[23–25]. Although the frequency range associated with local resonance is generally small, graded resonators can be coupled to considerably extend the frequency range by creating a rainbow effect[26,27]. Interesting tonotopic features can also be obtained in a solid structure with continuously varying geometrical properties as done by Dal Poggetto et al.[28], where the tonotopic effect is achieved without the use of locally resonant structures. In this case, the tonotopic effect was observed only in the out-of-plane displacements of the structure, due to the mainly flexural nature of the vibration modes. Here, we report on the design of a seashell-inspired resonator with a circular cross section exhibiting a tonotopic effect along two different polarization directions: one along the z-axis of the spiral, and the second in the coiling plane, perpendicular to the $z$-axis. The choice of a circular hollow cross section instead of a plate-like one is due to the similar resonant frequencies in the transverse and in-plane directions due to similar moments of inertia along perpendicular axes in the cross section. In the plate-like case, on the other hand, the moments of inertia with respect to the axes of the cross-section are considerably different concerning out-of-plane and in-plane vibration modes. Exploiting this design, the device not only displays tonotopy, but also sensitivity to the polarization of input or output pulse signals.

## II. METASENSOR DESIGN AND MODAL CHARACTERIZATION

### A. Surface parametrization

Similarly to what has been shown in Refs. [16,17] , several different shell structures can be generated with the same formula, as reported in Figure 1. The parametric expression for the central line, shown in Figure 1(a), which describes the 3D surface whose radius of the cross section $r(\theta)$ varies as a function of the curvilinear angle $\theta$, can be written as follows:

$$\begin{cases} x = R_0 \cos\theta \exp\left(k_R \dfrac{\theta}{\theta_{max}}\right), \\ y = R_0 \sin\theta \exp\left(k_R \dfrac{\theta}{\theta_{max}}\right), \\ z = R_a \theta \exp\left(k_a \dfrac{\theta}{\theta_{max}}\right). \end{cases} \tag{1}$$

where $\theta_{max} = 2\pi \cdot n_T$, and $n_T$ is the number of turns. $R_0$ is the initial curvature radius and $k_R$ is the reduction factor of the curvature radius, that is $R(\theta) = R_0 \exp\left(k_R \frac{\theta}{\theta_{max}}\right)$. Finally, $Ap_1 = R_a 2\pi \exp\left(\frac{k_a}{n_T}\right)$ is the axial pitch of the first turn i.e., the rate of change of the z coordinate with increasing $\theta$. The pitch decreases with increasing $\theta$ by means of the reduction factor $k_a$.The cross section of the seashell, centered around the central line, is usually described by an ellipse or a similar numerically-generated curve. Figure 1(c), (e) and (g) reports some examples of different shell species that can be generated, following Shojaei et al [17].

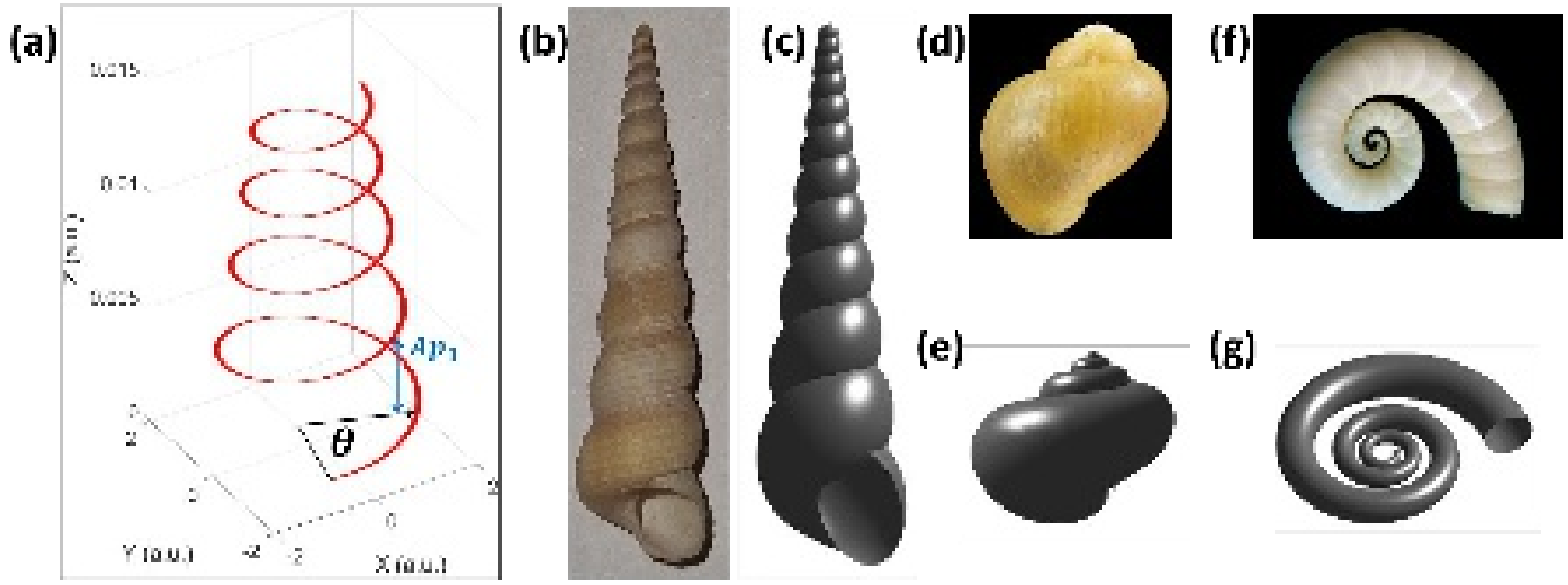

Figure 1 (a) Central line defining the spatial evolution of the shell-like surface along the $z$ axis; (b) *Turritellinella tricarinata; (c) Turritella*-like surface; (d) *Pomacea diffusa* (adapted from [29]) and (e) the associated surface. (f) and (g): same as for (e) and (f), but for the *Spirula spirula* shell.

### B. Modal analysis of different shell-inspired structures

In the following, we will consider three distinct shell-inspired structures that will mainly feature a different axial pitch, as shown in Figure 2(a): a *Turritella*-like structure, with the largest pitch, an intermediate one and, finally, a structure with zero axial pitch i.e., purely planar, similar to the *S. spirula* shell. Different Boundary Conditions (BCs), fixed or free, can be considered for the bottom cross section and for the apex. For simplicity, the cross section is taken as circular, and is expressed by $r(\theta) = r_g \exp\left(k_b \frac{\theta}{\theta_{max}}\right)$, where $r_g$ is the initial radius of the cross section, which also reduces as a function of the curvilinear angle via the reduction factor $k_b$. Moreover, we will also avoid intersections between different turns. This condition is achieved by imposing a minimum distance between the inner surface of a turn and the outer surface of the successive one. The vibrational modes of the structure are derived by Finite Element Analysis (FEA) by considering the surface as a structural shell element with both displacement and rotational degrees of freedom. The thickness is given by $\frac{r_g}{k_{th}}$, where $k_{th}$ is a constant factor that we usually choose equal to 10 or 15. The FEA is performed using the Shell interface of the Structural Mechanics module in COMSOL Multiphysics and more details are reported in the Appendix.

The mechanical properties of the material are chosen as those of a standard polymer used in 3D printing such as, for instance, Solflex SF650 (W2P Engineering GmbH), with Young's modulus $E = 2.5$ GPa, density $\rho = 1150\ \mathrm{kg/m^3}$ and Poisson's ratio $\nu = 0.33$. This is in view of a future experimental implementation, which will rely on 3-D printing as the most likely fabrication technique. An eigenfrequency study was performed and, for every eigenfrequency, $\omega_i$ and value of the curvilinear angle, $\theta_j$ the average of the absolute displacements was computed with respect to 10 points along the cross section, $\overline{U}(\omega_i, \theta_j) = \frac{\sum_{k=1}^{n} |\boldsymbol{u}_k(\omega_i, \theta_j)|}{n}$ with $n = 10$, and normalized to the maximum average obtained for all the curvilinear angles at that eigenfrequency, $\max\,[\overline{U}(\omega_j, \theta)]$, finally giving $U(\omega_i, \theta_j) = \frac{\overline{U}(\omega_i, \theta_j)}{\max\,[\overline{U}(\omega_i, \theta)]}$. The bottom right picture of Figure 2(a) shows the location of these points as red symbols along the cross section of a portion of the planar structure, for four different values of the curvilinear angle, $\theta$.

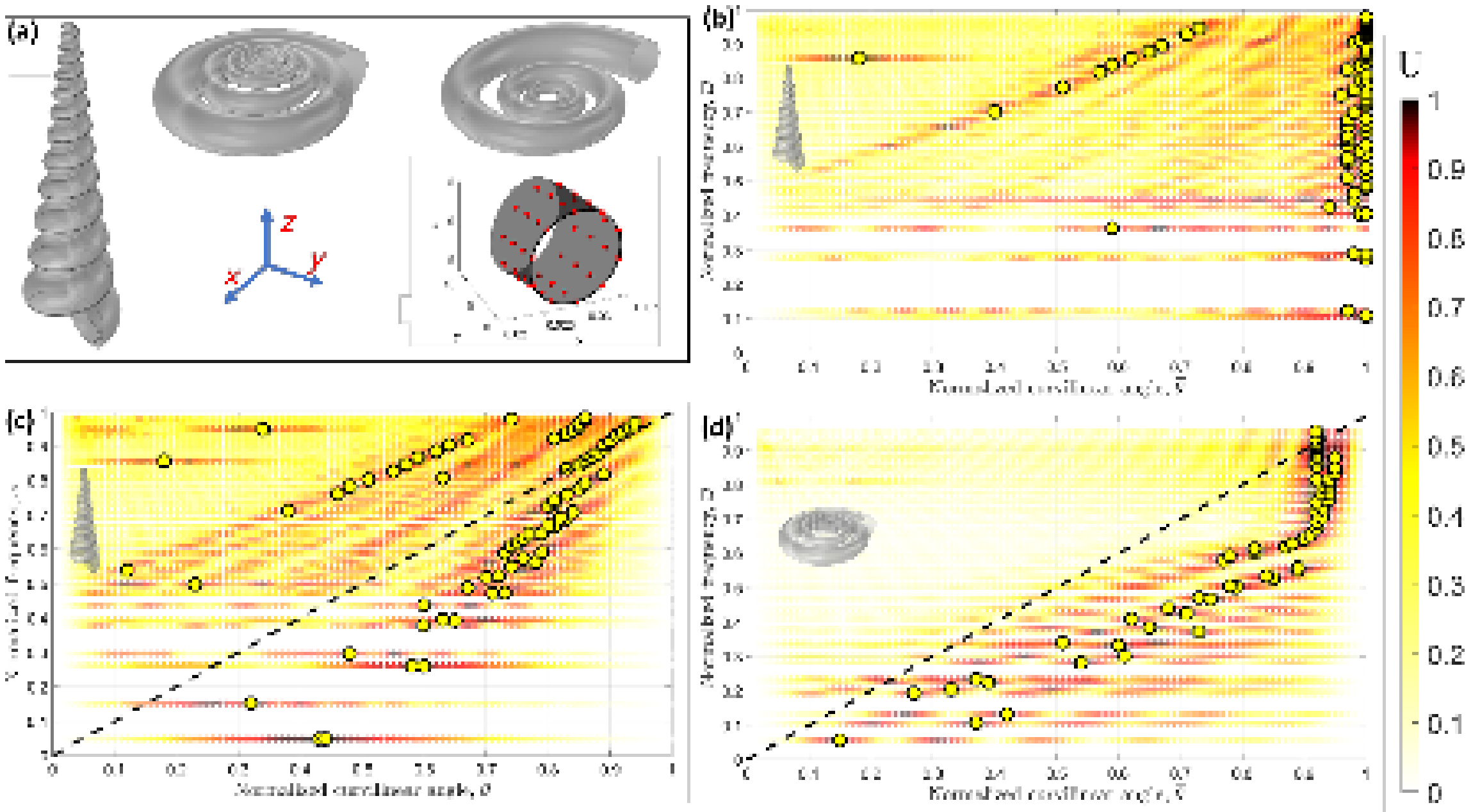

Figure 2 (a) The three analyzed structures with an example, in the lower right part of the Figure, of the geometrical locations used for calculating the displacements, indicated with red symbols. (b) Eigenfrequency study of the *Turritella*-like structure with fixed-free boundary conditions, that is, the base is fixed, and the apex is free. Colormap represents the displacements $U(\omega_i, \theta_j)$ and the yellow symbols the maximum $U$ for that eigenmode. (c) same as in (b) but with fixed-fixed BCs. (d) The axial pitch is reduced, and BCs are fixed-fixed.

Figure 2(b) and (c) report the results for the *Turritella*-like structure with the fixed-free and fixed-fixed BCs for the bottom and the apex, respectively. More information on the BCs is reported in the Appendix. The abscissae are the normalized curvilinear angle $\overline{\theta} = \theta/\theta_{max}$, the ordinates are the normalized logarithmic frequency $\overline{\omega} = \frac{\ln\left(\frac{\omega_i}{\omega_{min}}\right)}{\ln\left(\frac{\omega_{max}}{\omega_{min}}\right)}$, where $\omega_i$ is the angular frequency of the $i^{th}$ eigenmode and $\omega = 2\pi f$. We chose $f_{min} = 100$ Hz and $f_{max} = 10$ kHz. A maximum number of 100 eigenmodes is considered in the simulations, although considerably fewer occur in the chosen frequency range for the investigated structures. The color map represents the absolute displacements $U(\omega_i, \theta_j)$, while the yellow symbols correspond to the maxima for the corresponding eigenmodes in the $(\overline{\omega}, \overline{\theta})$ plane. As discussed in Liu et al[11], the energy concentrates at the apex i.e., close to $\overline{\theta} = 1$, possibly to protect the organism living inside the seashell against impacts. This can be seen in a more detailed way in the fixed-free configuration shown in Figure 2(b), where the maximum displacement is concentrated at the apex for most of the eigenmodes. At the same time, one family of eigenmodes shows a linear trend on the semi-logarithmic scale of the plot, which is reminiscent of the tonotopic behavior observed in the mammalian cochlea or in cochlea-inspired devices[28], where there is an approximately linear correlation between frequency and curvilinear angle. If we now change the BCs to fixed-fixed (Figure 2(c)), we notice an enhancement of this behavior, where several modes are now distributed linearly in the $(\overline{\omega}, \overline{\theta})$ plane. As it will be discussed later in more details in Figure 3 for another structure, by looking at the modal shapes of these eigenmodes, it is possible to notice that almost all the modes above the dashed line are predominantly polarized along the *z*-axis, while those below the line are dominant in the $xy$ plane (not shown in the Figure). The black dashed line represents an ideal tonotopic behavior described by the $\overline{\omega} = \overline{\theta}$ line, while the isolated maxima in the left part of the graph are circumferential modes. An example of this type of modes is shown in Figure 3 for a different structure. By keeping the same BCs but reducing the axial pitch along $z$ so that the structure is "flattened", we can now observe in Figure 2(d) that the tonotopic behavior is improved because there are fewer parallel linear branches. Figure 3 illustrates the case when the axial pitch is reduced to zero, i.e., the structure is planospiral: the same type of behavior occurs, with the maxima distributed along two almost parallel lines and, in the top left corner, it is possible to notice the presence of some circumferential modes. For this structure, the different BCs, fixed-fixed of fixed-free, do not give rise to considerable differences in the tonotopic behavior, apart from the fact the latter BCs give rise to the presence of a larger number of eigenmodes in a linear trend, as shown in Figure 3.

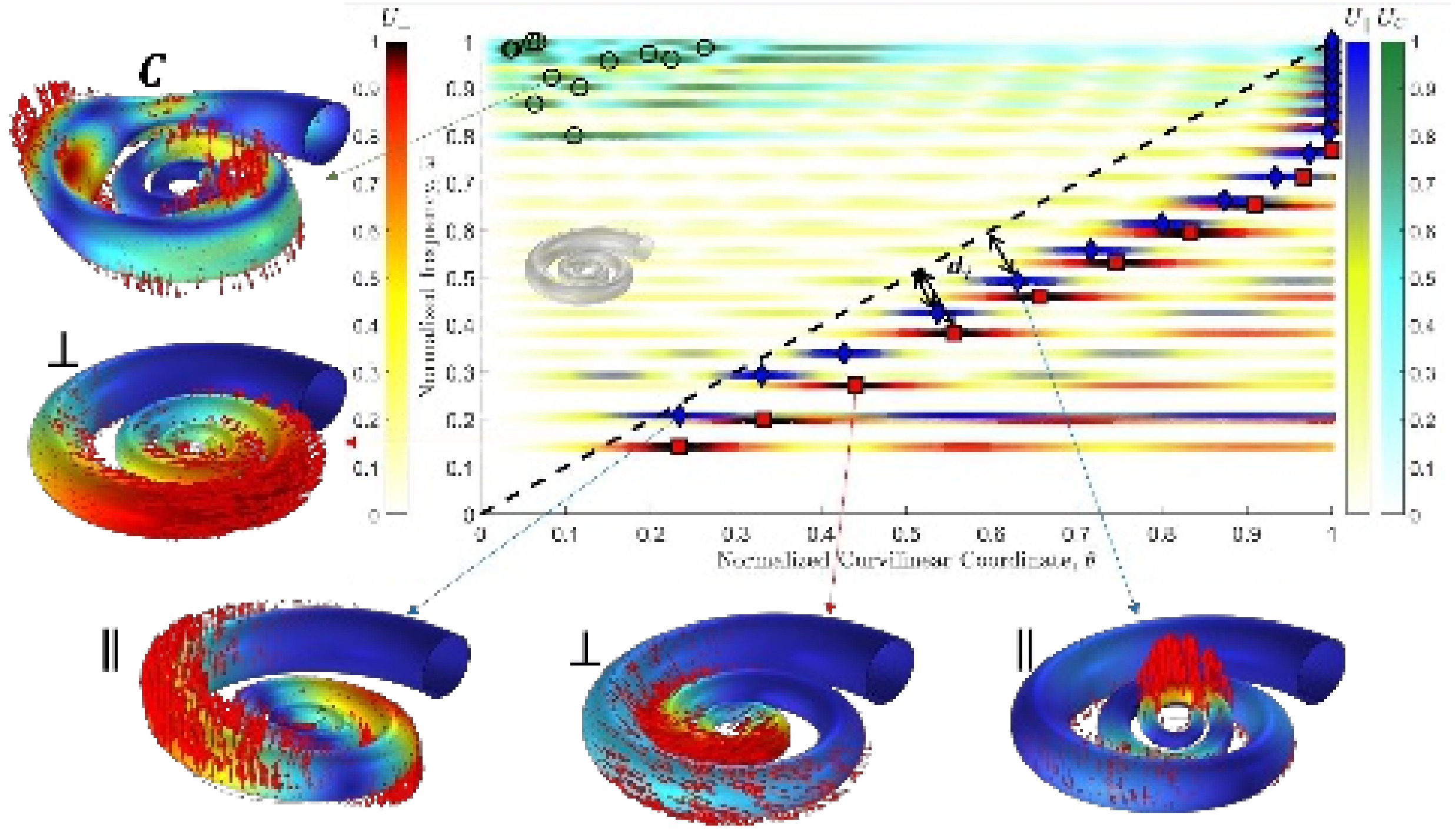

Figure 3: Eigenfrequency study of the structure when the axial pitch is reduced to zero i.e., the structure is planospiral, and the BCs are fixed-free. Main panel: color bars represent the normalized displacements as a function of the curvilinear angle and eigenfrequency. The different color maps represent the different types of modes, circumferential, predominantly-axial or predominantly-planar. Blue diamonds and red squares represent the maximum axial and planar displacements, respectively, while circles refer to circumferential modes. Black dashed line represents the ideal tonotopic line. The other sketches report some examples of modal shapes for the three types of modes. Red arrows represent the displacement vector.

It is also possible to analyze the modal shapes of the different eigenmodes. The upper left Figure reports the modal shape of a circumferential mode (indicated as "C"). This can be seen by the deformation of the cross section at various positions along the structure. It is worth recalling here that the initial part of the structure (larger radius of the cross section) is fixed. We can distinguish two other types of modes in this structure. The first is characterized by displacements mainly parallel to the $z$ axis, and the other by displacements that are dominant in the coiling plane. By taking as a reference the $z$ axis, we will call parallel (∥) the first type, and perpendicular (⊥) the second.

Figure 3 shows two modal shapes for each of these two families, as indicated by the corresponding labels. The different vibration direction for the two different cases is clear. If we call $u, v, w$ the components of the displacement along the $x, y$ and $z$ axis, respectively, it is then possible, according to the previous discussion, to calculate the parallel and perpendicular displacements. While the displacement along $z$ is simply $u_{||} = w$, the perpendicular one can be regarded as $u_{\perp} = \sqrt{u^2 + v^2}$. Then, similarly to what has been done in the case of the total displacement, it is possible to calculate the normalized displacements $U_{\parallel}(\omega_i, \theta_j)$, and $U_{\perp}(\omega_i, \theta_j)$ and make the same considerations as in the case of the eigenfrequency study for the total displacement. It is possible to see that all modes display either a dominant vibration along the $z$-axis or in the coiling plane. The absolute displacements for the three different types of eigenmodes are displayed in Figure 3 with three different dedicated color maps where the corresponding maxima are represented as blue diamonds and red squares for the parallel and perpendicular modes, respectively. Maxima related to circumferential modes are indicated by empty circles. Interestingly, both axial and planar modes partially follow a tonotopic trend. Thus, a distance $d_i$ in the $(\overline{\theta}, \overline{\omega})$ plane from a given maximum displacement and the desired ideal tonotopic line can be defined, as shown by the black dashed lines.

## III. OPTIMIZATION AND SENSING APPLICATIONS

### A. Optimization

The geometrical parameters describing the structure featuring the behavior shown in Figure 3 can be optimized to achieve the behavior indicated by the $\overline{\omega} = \overline{\theta}$ line. The procedure can be performed by optimizing at the same time the displacements along the two directions by minimizing the sum of the distances $d_i$ from the $\overline{\omega} = \overline{\theta}$ line, in a similar way as done in ref [28], where the focus was on one type of excitation only. A metric function Γ is defined as $\Gamma = \frac{\sum_i^n d_i^2}{n}$ where $n$ is the number of considered eigenmodes. Since the structure is planar, $k_a = 0$. Furthermore, to make all the structures comparable, we fix the number of turns to $n_T = 3$ and the mass (or, equivalently, the volume) of the structure to be optimized. Finally, to avoid intersections, that would considerably modify the vibrational behavior of the structure, we add an additional constraint that sets the minimum distance between two successive turns to $\Delta_R^{min} = 10^{-4}$ m. Thus, the parameters to be optimized are only the two reduction factors, $k_R$ for the curvature radius, $R(\theta)$ and $k_b$ for the radius of the cross section, $r(\theta)$. Another constraint is applied to $k_b$ to avoid that the radius of the cross section at the end of the structure, $r(\theta = 2\pi n_T)$ becomes smaller than the thickness of the structure. The procedure is implemented by using a Matlab sequential quadratic programming algorithm to solve the constrained optimization problem. Figure 4(a) shows a more detailed diagram of the steps followed in the optimization process.

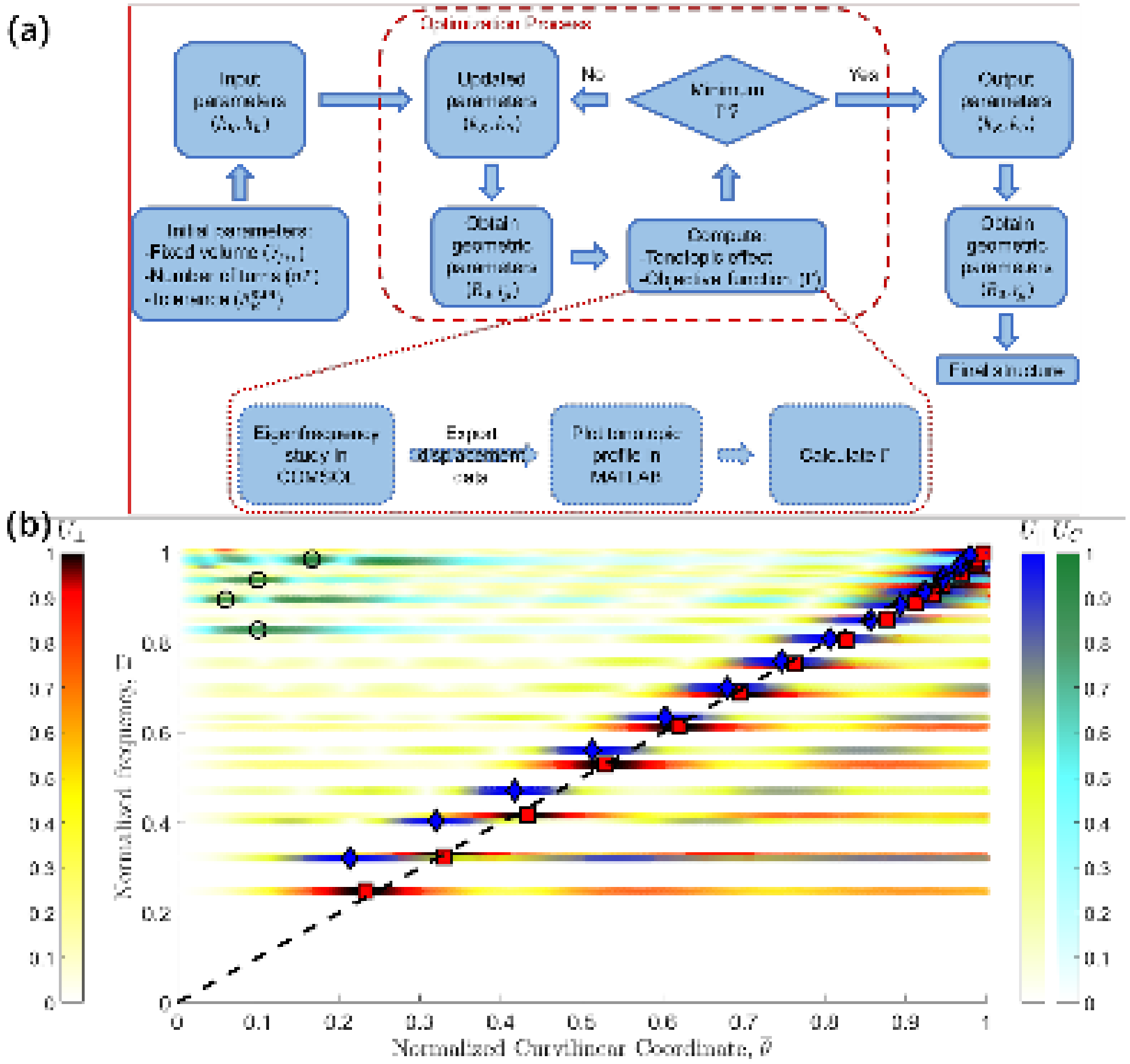

Figure 4: (a) Flow chart illustrating the steps followed to optimize the tonotopic structure. (b) color plot representing the normalized displacements as a function of the curvilinear angle and eigenfrequency for the optimized structure. The three types of modes, predominantly-axial, predominantly-planar and circumferential are represented by different color maps. Blue diamonds and red squares represent the maximum axial and perpendicular displacements, respectively and circles indicate circumferential modes.

The parameters of the optimized structure, reported in Figure 5(a), are $k_{R,opt} = -2.97$ and $k_{b,opt} = -2.57$. Then, to avoid intersections between different turns and to maintain the volume fixed, we also obtain $r_{g,opt} = 6.9 \times 10^{-3}$ m and $R_{0,opt} = 2.9 \times 10^{-2}$ m. Figure 4(b) reports the result of the optimization process which maximizes the tonotopy both along the axial and perpendicular direction at the same time. It is possible to notice that the maximum displacement of the axial and perpendicular modes follows a tonotopic trend for almost two decades for both types of excitations, displayed with two different color maps.

### B. Tonotopy and sensitivity to polarization

We now focus on evaluating the possibility of exploiting the tonotopic properties of the designed structures for vibration sensing applications. This can be verified by means of time-domain simulations which can reproduce configurations close to future experimental conditions. These are performed on the optimized structure previously obtained described by the tonotopic profile reported in Figure 4(b). First, it is necessary to demonstrate that the frequency of a given signal can be determined by observing the curvilinear angle of the maximum displacement and, from that, to check that the predicted frequency matches with the one emerging from propagation simulations.

For the time-domain simulations, a short segment of a straight hollow cylinder of the same material as the resonator and with the same radius and thickness as the initial cross section i.e., at $\theta = 0$, is attached to the device, as highlighted in red in Figure 5(a). The edge of the other end is fixed, to reproduce the boundary conditions used in the optimization process. The input signal is then applied on the whole surface of the attachment. An input polarization angle, $\varphi$ can be defined with respect to the $xz$ plane of Figure 5(a): $\varphi = 0$ refers to an input signal parallel to $x$, while $\varphi = 90°$ parallel to the $z$-axis. In the simulations, the input polarization angle is initially set to $\varphi = 45°$. The input signal is a Gaussian-modulated sine wave, and the width of the Gaussian window is 10 sine periods. The total duration of the input is 20 periods of the sine wave. Panels (b) and (c) of Figure 5 show the time-domain signal and the corresponding spectrum in terms of fundamental period and frequency, respectively. More technical details regarding the FEA in the time domain are reported in the Appendix.

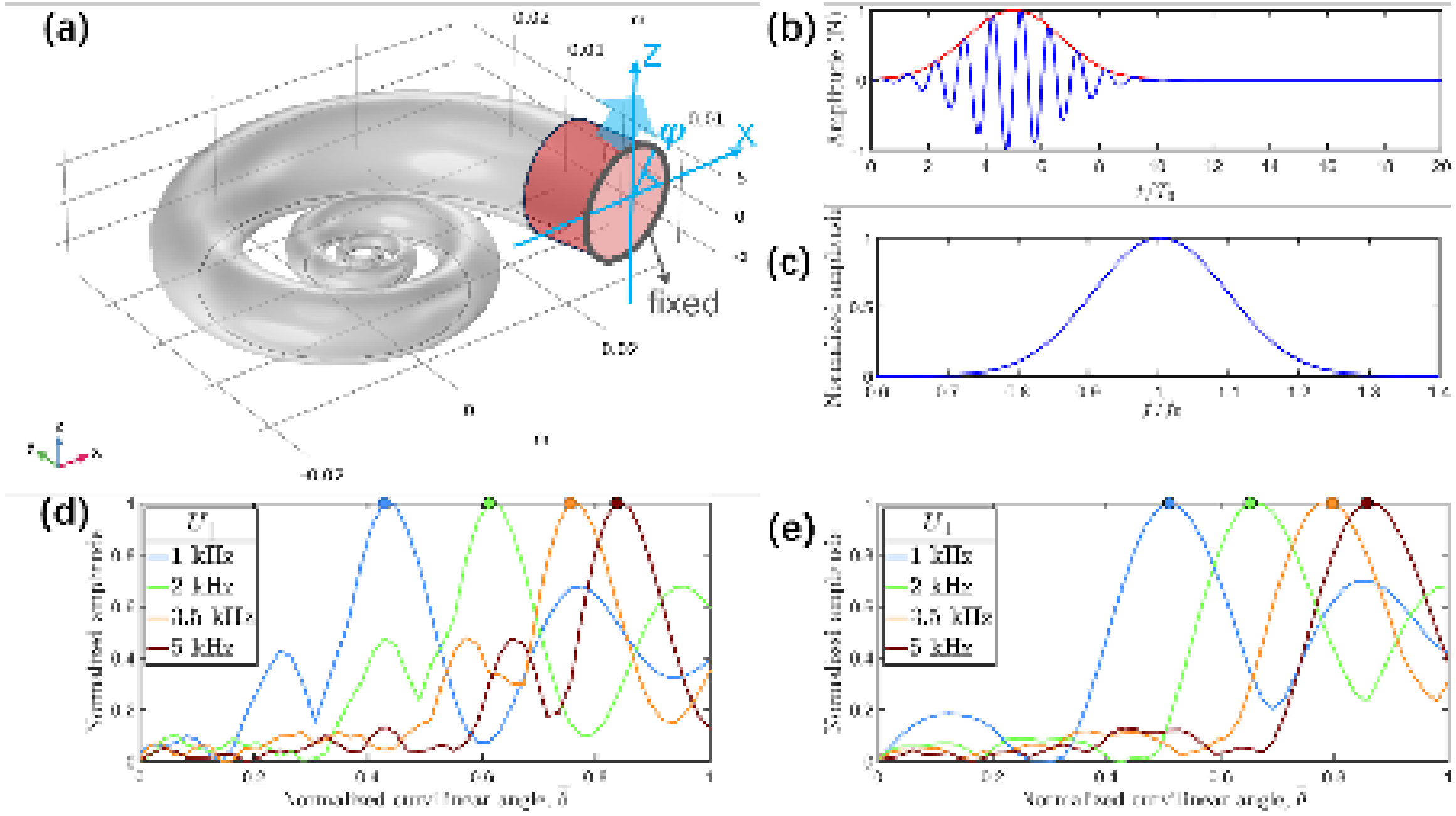

Figure 5: Time-domain simulations: (a) optimized structure and loading configuration with variable input polarization angle, $\varphi$. The arrow represents a polarization angle $\varphi = 45°$. (b) Gaussian-modulated input time-domain signal used for the simulations and (c) corresponding frequency content. (d) Normalized maximum axial displacement, $U_{\parallel}$ as a function of the normalized curvilinear angle for input signals at different frequencies. (e) same as (d), but for the planar displacement, $U_{\perp}$.

Four different central frequencies for the pulse are tested, namely $f_1 = 1$ kHz, $f_2 = 2$ kHz, $f_3 = 3.5$ kHz, $f_4 = 5$ kHz. For each of the two selected directions of displacement, axial and perpendicular, and for each value of the curvilinear angle $\theta$, the maximum of the signal recorded in time is plotted in Figure 5(d) and (e), respectively. The signals are usually measured in 50 $\theta$ points either along a line tangent to the outer surface of the structure and lying in the $xy$ plane (see in Figure 5(a)) or on the line tangent the top of the surface. For each frequency, a clear peak emerges for the recorded signal at a precise curvilinear angle, confirming the expected tonotopic behavior. The reported curves are normalized to the overall maximum of each signal, which also identifies the relevant normalized curvilinear angle $\overline{\theta}$ associated to the input frequency. Notice that the width of the maxima is narrower in the case of $z$-polarized signals, indicating a polarization direction along which measurements provide greater precision. These results are summarized in Figure 6(a), to highlight the overall behavior. The symbols used in Figure 6(a) correspond to the curvilinear angle of the signal maximum at the corresponding frequency, as determined in Figure 5(d) and (e), while the blue and red lines, which serve as a reference, correspond to the lines connecting the maximum displacement of the eigenmodes of the optimized structure that were previously displayed as symbols in Figure 4(b) for the two different polarizations, axial and planar. For the sake of clarity, the two lines have been slightly smoothed. Symbols relative to time-domain simulations are reported with error bars to quantify the angular uncertainty due to the finite number of detection angles in simulations. Results indicate excellent agreement between frequency domain and time domain simulations. Further, analysis of the time-domain signals, which are indicative of potential measurements in an experimental setting, allows the determination of the input frequency with remarkable precision, by detecting the curvilinear angle of maximum displacement.

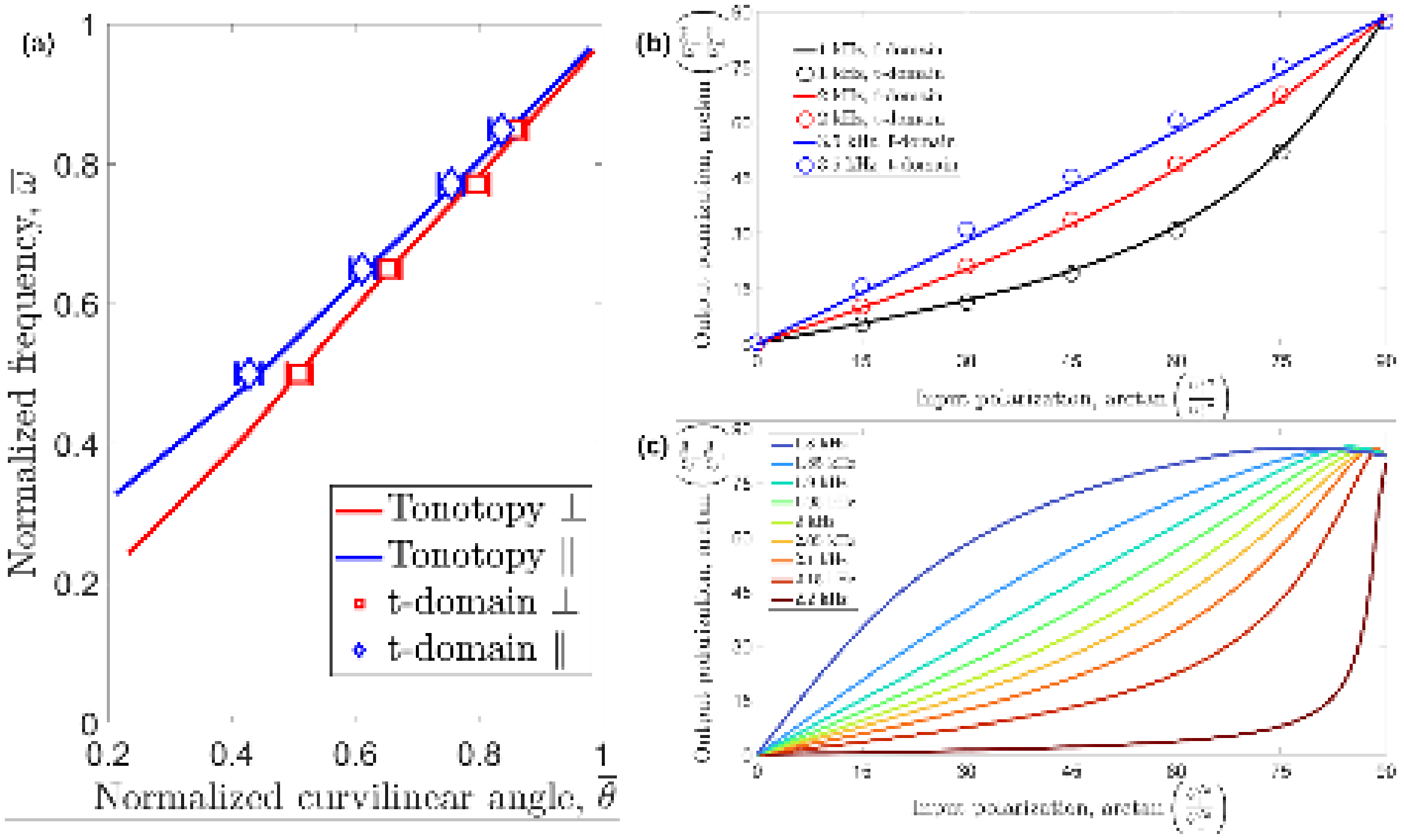

Figure 6: (a) Blue and red lines: tonotopically optimized eigenmodes for the axial and planar displacements, respectively. Symbols: time-domain results at different frequencies and for polarization $\varphi = 45°$. (b) Output polarization vs. Input polarization for different frequencies: lines correspond to frequency domain simulations for 90 different polarization angles while symbols correspond to time-domain analysis for 7 different polarization angles. Black: 1 kHz. Red: 2 kHz. Blue: 3.5 kHz. (c) Output polarization vs input polarization for some frequencies in the $1.8 - 2.2$ kHz range, showing a strong polarization variation with the input frequency.

Results also highlight the capability of the optimized device to simultaneously sense two polarization directions. It is therefore interesting to see whether by analyzing the vibration of the resonator, it is also possible to determine the polarization of the input signal or, on the other hand, if it is possible, by suitably tuning the input polarization and frequency, to obtain a desired polarization output. Moreover, given the input signal polarization and frequency, this would also allow to estimate the full-field displacements by measuring only a single displacement component. Additional simulations are thus performed in the time domain by varying the input signal polarization, $\varphi$, and by the analysis of the detected output signal. Three different frequencies are tested i.e., $f_1 = 1$ kHz , $f_2 = 2$ kHz and $f_3 = 3.5$ kHz and seven different input signal polarization: $\varphi = 0°, 15°, 30°, 45°, 60°, 75°$ and $90°$. The output signals are detected in the same locations as for the $\varphi = 45°$ case, reported in Figure 6(a). Figure 6(b) reports the output polarization i.e., the arctangent of the ratio of the axial displacements over the planar one $U_{\parallel}^{Out}/U_{\perp}^{Out}$ for the three selected frequencies. The results are reported as a function of the input polarization and the displacements are measured at the tonotopic curvilinear angle related to the corresponding frequency. Symbols represent the time-domain results, obtained with the same procedure used for the $\varphi = 45°$ case and discussed above. Lines are the same quantity but obtained with a frequency domain simulation as a function of 90 different input angles. The agreement is remarkable, all the more since, to accurately reproduce realistic experimental conditions, a gaussian pulse is adopted as input signal with a rather wide frequency spectrum, as seen in Figure 5(c). Indeed, the device is excited with multiple frequencies, but the dominant frequency is sufficient, in frequency domain analysis, to estimate the input polarization direction. Thus, this type of analysis can reliably predict the response of the device at different frequencies and different polarizations and can be used, for example, as a reference to determine the polarization at any frequency. The localization of the curvilinear angle of the maximum displacement, as explained in the previous section, allows to estimate the input frequency, exploiting tonotopy. In addition, by measuring output polarization, it is also possible to obtain a good estimation of the input one. Furthermore, apart from the 3.5 kHz case, the output polarization is different from the input one. This effect is more evident, the more the behavior at a given frequency deviates from the bisector line. If this effect can be controlled, the device could be used as a polarizer, that is, any desired output polarization can ideally be achieved, as a function of the input. As an example, Figure 6(c) shows that, in the 1.8-2.2 kHz range, a strong tuning of the output polarization can be obtained as a function of the input frequency for fixed input polarization. For an input signal polarization $\varphi = 75°$, the output one becomes $\varphi = 84.5°$ at 1.8 kHz, and $\varphi = 7.8°$ at 2.2 kHz, i.e., it passes from an almost axial polarization to a planar one. As discussed in the next section, this feature can be exploited in applications, e.g. when multiple harmonic components are present in an input signal, such as in Non-Destructive Testing (NDT) techniques[30]. These harmonics can appear due to the presence of defects that give rise to nonlinearities. Polarization control can therefore be used in conjunction with tonotopic spatial frequency mapping to better discriminate their presence, despite their low amplitude.

## IV. DISCUSSION AND CONCLUSIONS

In this work, we have investigated the potentially non-trivial dynamic properties emerging from bioinspired shell-like structures, focusing on the intriguing characteristic of tonotopy that naturally emerges in geometric structures where curvature and grading act as distinctive elements. Specifically, we have focused on a three-dimensional spiral structure that is symmetrical with respect to the $xy$-plane, in which grading is introduced by varying the cross-sectional shape and the curvature radius along the length of the resonator. We have shown that this type of 3D geometry allows to simultaneously achieve tonotopic features on two types of modes, with orthogonal polarization to each other, i.e. prevalently planar or axial. We have described a geometrical optimization process that allows to achieve the most effective spatial distribution of vibrational modes for both polarizations at the same time, resulting in a planospiral metasensor with tonotopic properties along both the z-axis and in the xy-plane. Both frequency-domain and time-domain simulations employing realistic multifrequency pulsed signals have highlighted the potential of the metasensor to not only spatially separate different frequencies, but also to control their polarization, univocally determining the output vibration direction as a function of the input one, or vice versa.

Given the bioinspired nature of the design, this might have implications from a bio-evolutionary standpoint. The functionality of the different shell designs discussed in this paper might thus be linked to their impact resistance and elastic wave transmission properties. For example, it has been clearly shown for organisms such as the Mantis Shrimp that microstructure directionality plays an important role in deviating and damping impact energy to avoid fracture[31]. It might also be that macrostructure is involved in this process, by efficiently converting longitudinal vibrations to shear or vice versa along the axis of the shell.

From a practical application standpoint, the considered structure can be exploited for signal analysis and advanced sensing of elastic waves. In particular, this device acts as an elastic analogue of the optical prism, spatially separating the frequency content of impinging propagating signals. Thus, its tonotopic nature can be exploited for the optimization of the signal-to-noise ratio of specific frequencies by identifying points of maximal vibration for the signal detection. This can be important in devices for speech recognition[32] or artificial cochlear implants[33,34], where the separation of time signals into different frequency-dependent channels is a key point in achieving accurate and efficient processing of auditory information. Another sector of potential interest is that of Internet of Things (IoT), in which power consumption is of paramount importance. A further possible application of the device is in non-destructive testing (NDT) sensors, where low-amplitude higher harmonics emerging from defects/nonlinearities need to be discriminated from fundamental frequencies[30]. This can lead to a more efficient analysis, circumventing the need for high-energy processes typical of conventional sensors. Integrating tonotopy into a sensor design could thus be an original contribution to NDT sensor technology for the monitoring of structural integrity. We could consider, for instance, a scenario in which a smart sensor is embedded in a structural member and functions as low-power real-time acoustic emission (AE) monitoring device. In general, it can be used as a passive frequency separator for real-time analysis of the harmonic content of a signal, working in a similar way to elastic demultiplexers[35].

Moreover, the distinctive feature of our device is the extension of tonotopic properties onto two perpendicular planes. This characteristic appears to be unique, and is inherently linked to tonotopic structure, where energy is locally focused on different points across the entire structure, depending on its frequency content. The first advantage of such a structure, if employed as a tonotopic sensor, is that it allows to determine full-field displacements by measuring only a single polarization component. Other possible applications directly emerging from this work are a frequency-dependent polarizer or an analyzer of input signal polarization. This can be particularly useful in NDT, specifically in anisotropic materials like composites, where the acoustic radiation pattern is highly directional. Indeed, the capability of analyzing wave polarization could be an additional feature that allows to distinguish between different failure modes (e.g. cracks or delaminations), based on the coupled analysis of both frequency[36] and directionality[37]. Another possible application is in the field of speech recognition and signal processing, with the prospect of creating an artificial cochlea. In particular, the ability to distinguish different polarizations in the elastic excitation of the proposed sensor could be exploited in conjunction with MEMS directional microphones[38] for the transduction of an acoustic signal into a mechanical one that also contains information about the sound directionality.

In general, the adoption of a bioinspired approach in the development of acoustic and elastic metamaterials and devices not only enhances their performance, but also unlocks innovative design possibilities, paving the way for sustainable and efficient solutions in diverse applications.

## ACKNOWLEDGEMENTS

All authors acknowledge the European Commission under the FET Open ''Boheme'', Grant No. 863179. Y.L. acknowledges the China Scholarship Council, Grant No. 202108430020. We are thankful to Marco Scalerandi, M. Faghih Shojaei and H.Rajabi for useful discussions.

## APPENDIX: FINITE ELEMENTS ANALYSIS DETAILS

### A. Eigenfrequency study

Finite element simulations are performed on models considering structural shell elements, since the tonotopic effects mainly depend on flexural modes. The representative matrix equation, in this case, is:

$$\boldsymbol{M}\frac{\partial^2 \boldsymbol{u}}{\partial t^2} + \boldsymbol{C}\frac{\partial \boldsymbol{u}}{\partial t} + \boldsymbol{K}\boldsymbol{u} = \boldsymbol{f}(t)$$

where $\boldsymbol{M}$ is the mass matrix, $\boldsymbol{C}$ is the damping matrix, $\boldsymbol{K}$ is the stiffness matrix, and $\boldsymbol{f}(t)$ is the external force. Both finite element matrices and external force vectors are obtained through the usual assembly processes considered in a finite element framework and implemented in COMSOL Multiphysics.

Since in the considered modal analyses the system vibrates freely with no damping, the matrix equation reduces to:

$$(-\omega^2 \boldsymbol{M} + \boldsymbol{K})\widetilde{\boldsymbol{u}} e^{i\omega t} = 0,$$

where $\omega^2$ represents the resonance frequencies of interest and $\widetilde{\boldsymbol{u}}$ represents the vibration mode shape.

The eigenvalues of the previous equation can thus be found by solving:

$$det(-\omega^2 \boldsymbol{M} + \boldsymbol{K}) = 0$$

### B. Boundary conditions

For the fixed BC, the displacement is set to zero, i.e. $\boldsymbol{u} = 0$. The fixed-fixed and fixed-free BCs correspond to $\boldsymbol{u}_{\theta=0} = \boldsymbol{u}_{\theta=2\pi\cdot n_T} = 0$ and $\boldsymbol{u}_{\theta=0} = 0$, respectively.

### C. Time-domain simulations

The relationship between the mesh size and time step, determined by the Courant-Friedrichs-Lewy (CFL) condition[39], is used for the time-domain simulations. This is related to the Courant number:

$$CFL = \frac{c\Delta t}{h},$$

where $c$ is the wave velocity, $\Delta t$ the time step and $h$ the mesh element size. The time step for the simulations is determined as:

$$\Delta t = \frac{h_{max} CFL}{c} = \frac{CFL}{f_{max} N},$$

where $h_{max}$ is the maximum element size, and $N$ represents the minimum number of mesh elements per wavelength (in principle, the maximum element should be less than $\lambda_{min}/5$, where $\lambda_{min}$ is the minimum wavelength). With the default 2nd-order, quadratic discretization setting adopted in the shell interface in COMSOL, the CFL number should be less than 0.2, and a value of 0.1 proves to be nearly optimal. In our study, the CFL is therefore set as 0.1, $N$ to 6, so that $\Delta t \approx \frac{1}{60 f_{max}}$.

## REFERENCES

[1] U.G.K. Wegst, and M.F. Ashby, "The mechanical efficiency of natural materials," Philos. Mag. **84**(21), (2004).

[2] P. Fratzl, and R. Weinkamer, "Nature's hierarchical materials," Prog. Mater. Sci. **52**(8), (2007).

[3] Z. Gan, M.D. Turner, and M. Gu, "Biomimetic gyroid nanostructures exceeding their natural origins," Sci. Adv. **2**(5), (2016).

[4] S. Tadepalli, J.M. Slocik, M.K. Gupta, R.R. Naik, and S. Singamaneni, "Bio-Optics and Bio-Inspired Optical Materials," Chem. Rev. **117**(20), (2017).

[5] S. Dou, H. Xu, J. Zhao, K. Zhang, N. Li, Y. Lin, L. Pan, and Y. Li, "Bioinspired Microstructured Materials for Optical and Thermal Regulation," Adv. Mater. **33**(6), (2021).

[6] V.F. Dal Poggetto, "Bioinspired acoustic metamaterials: From natural designs to optimized structures," Front. Mater. **10**, (2023).

[7] U.G.K. Wegst, H. Bai, E. Saiz, A.P. Tomsia, and R.O. Ritchie, "Bioinspired structural materials," Nat. Mater. **14**(1), (2015).

[8] B.S. Lazarus, A. Velasco-Hogan, T. Gómez-del Río, M.A. Meyers, and I. Jasiuk, "A review of impact resistant biological and bioinspired materials and structures," J. Mater. Res. Technol. **9**(6), (2020).

[9] F. Bosia, V. Dal Poggetto, A.S. Gliozzi, G. Greco, M. Lott, M. Miniaci, F. Ongaro, M. Onotato, S.F. Seyyedizadeh, M. Tortello, and N.M. Pugno, "Optimized structures for vibration attenuation and sound control in nature: A review," Matter **5**, 3311–3340 (2022).

[10] Z. Shen, T.R. Neil, D. Robert, B.W. Drinkwater, and M.W. Holderied, "Biomechanics of a moth scale at ultrasonic frequencies," Proc. Natl. Acad. Sci. U. S. A. **115**(48), (2018).

[11] Y. Liu, M. Lott, S.F. Seyyedizadeh, I. Corvaglia, G. Greco, V.F. Dal Poggetto, A.S. Gliozzi, R. Mussat Sartor, N. Nurra, C. Vitale-Brovarone, N.M. Pugno, F. Bosia, and M. Tortello, "Multiscale static and dynamic mechanical study of the Turritella terebra and Turritellinella tricarinata seashells," J. R. Soc. Interface **20**(205), (2023).

[12] M. Yourdkhani, D. Pasini, and F. Barthelat, "The hierarchical structure of seashells optimized to resist mechanical threats," WIT Trans. Ecol. Environ. **138**, (2010).

[13] E. Setyowati, G. Hardiman, Purwanto, and M.A. Budihardjo, "On the role of acoustical improvement and surface morphology of seashell composite panel for interior applications in buildings," Buildings **9**(3), (2019).

[14] H. Li, J. Shen, Q. Wei, and X. Li, "Dynamic self-strengthening of a bio-nanostructured armor — conch shell," Mater. Sci. Eng. C **103**, (2019).

[15] Z. Yin, F. Hannard, and F. Barthelat, "Impact-resistant nacre-like transparent materials," Science (80-. ). **364**(6447), (2019).

[16] J.L. Pappas, and D.J. Miller, "A Generalized Approach to the Modeling and Analysis of 3D Surface Morphology in Organisms," PLoS One **8**(10), (2013).

[17] M.F. Shojaei, V. Mohammadi, H. Rajabi, and A. Darvizeh, "Experimental analysis and numerical modeling of mollusk shells as a three dimensional integrated volume," J. Mech. Behav. Biomed. Mater. **16**(1), (2012).

[18] Fay R.R and Dallos P., *The Cochlea* (Springer Science and Business Media, Berlin, Germany, 2012).

[19] L. Robles, and M.A. Ruggero, "Mechanics of the mammalian cochlea," Physiol. Rev. **81**(3), (2001).

[20] J. Lighthill, "Biomechanics of hearing sensitivity," J. Vib. Acoust. Trans. ASME **113**(1), (1991).

[21] M. LeMasurier, and P.G. Gillespie, "Hair-cell mechanotransduction and cochlear amplification," Neuron **48**(3), (2005).

[22] T. Reichenbach, and A.J. Hudspeth, "The physics of hearing: Fluid mechanics and the active process of the inner ear," Reports Prog. Phys. **77**(7), (2014).

[23] F. Ma, J.H. Wu, M. Huang, G. Fu, and C. Bai, "Cochlear bionic acoustic metamaterials," Appl. Phys. Lett. **105**(21), (2014).

[24] F. Ma, J.H. Wu, M. Huang, and S. Zhang, "Cochlear outer hair cell bio-inspired metamaterial with negative effective parameters," Appl. Phys. A Mater. Sci. Process. **122**(5), (2016).

[25] M. Rupin, G. Lerosey, J. De Rosny, and F. Lemoult, "Mimicking the cochlea with an active acoustic metamaterial," New J. Phys. **21**(9), (2019).

[26] J. Zhu, Y. Chen, X. Zhu, F.J. Garcia-Vidal, X. Yin, W. Zhang, and X. Zhang, "Acoustic rainbow trapping," Sci. Rep. **3**(1), 1728 (2013).

[27] L. Zhao, and S. Zhou, "Compact acoustic rainbow trapping in a bioinspired spiral array of graded locally resonant metamaterials," Sensors (Switzerland) **19**(4), (2019).

[28] V.F. Dal Poggetto, F. Bosia, D. Urban, P.H. Beoletto, J. Torgersen, N.M. Pugno, and A.S. Gliozzi, "Cochlea-inspired tonotopic resonators," Mater. Des. **227**, (2023).

[29] H. Zell, "Own work, CC BY-SA 3.0 <https://creativecommons.org/licenses/by-sa/3.0>, via Wikimedia Commonsle," (n.d.).

[30] M. Miniaci, A.S. Gliozzi, B. Morvan, A. Krushynska, F. Bosia, M. Scalerandi, and N.M. Pugno, "Proof of Concept for an Ultrasensitive Technique to Detect and Localize Sources of Elastic Nonlinearity Using Phononic Crystals," Phys. Rev. Lett. **118**(21), (2017).

[31] J.C. Weaver, G.W. Milliron, A. Miserez, K. Evans-Lutterodt, S. Herrera, I. Gallana, W.J. Mershon, B. Swanson, P. Zavattieri, E. DiMasi, and D. Kisailus, "The stomatopod dactyl club: A formidable damage-tolerant biological

hammer," Science (80-. ). **336**(6086), (2012).

[32] T. Dubček, D. Moreno-Garcia, T. Haag, P. Omidvar, H.R. Thomsen, T.S. Becker, L. Gebraad, C. Bärlocher, F. Andersson, S.D. Huber, D.-J. van Manen, L.G. Villanueva, J.O.A. Robertsson, and M. Serra-Garcia, "In-Sensor Passive Speech Classification with Phononic Metamaterials," Adv. Funct. Mater. **n/a**(n/a), 2311877 (2024).

[33] J. Jang, J.H. Jang, and H. Choi, "Biomimetic Artificial Basilar Membranes for Next-Generation Cochlear Implants," Adv. Healthc. Mater. **6**(21), (2017).

[34] H. Tang, S. Zhang, Y. Tian, T. Kang, C. Zhou, S. Yang, Y. Liu, X. Liu, Q. Chen, H. Xiao, W. Chen, and J. Zang, "Bioinspired Soft Elastic Metamaterials for Reconstruction of Natural Hearing," Adv. Sci. **10**(20), (2023).

[35] O.R. Bilal, C.H. Yee, J. Rys, C. Schumacher, and C. Daraio, "Experimental realization of phonon demultiplexing in three-dimensions," Appl. Phys. Lett. **118**(9), (2021).

[36] M.G.R. Sause, T. Müller, A. Horoschenkoff, and S. Horn, "Quantification of failure mechanisms in mode-I loading of fiber reinforced plastics utilizing acoustic emission analysis," Compos. Sci. Technol. **72**(2), (2012).

[37] N. Ghadarah, and D. Ayre, "A Review on Acoustic Emission Testing for Structural Health Monitoring of Polymer-Based Composites," Sensors **23**(15), (2023).

[38] A. Ishfaque, and B. Kim, "Fly Ormia Ochracea Inspired MEMS Directional Microphone: A Review," IEEE Sens. J. **18**(5), (2018).

[39] R. Courant, K. Friedrichs, and H. Lewy, "On the Partial Difference Equations of Mathematical Physics," IBM J. Res. Dev. **11**(2), (2010).